\newcommand{\tw}{t_{\rm w}}
\newcommand{\teff}{t_{\rm eff}}

\newcommand{\Leff}{L_{\rm eff}}

\newcommand{\Lovlp}{L_{\rm \Delta T}}

\newcommand{\AgMn}{{Ag(11 at\% Mn) }}

\newcommand \be {\begin{equation}}
\newcommand \ee {\end{equation}}

\newcommand{\eq}[1]{Eq.~(\ref{#1})}
\newcommand{\kb}{k_{\rm B}}

\documentclass[aps,prl,twocolumn,superscriptaddress]{revtex4}

\usepackage{graphicx}% Include figure files

\begin{document}

\title{Symmetrical Temperature-Chaos Effect 
with Positive and Negative Temperature
Shifts in a Spin Glass}
\author{
P. E. J{\"o}nsson
}
\affiliation{
Department of Materials Science, Uppsala University, 
Box 534, SE-751 21 Uppsala, Sweden 
}
\author{
H. Yoshino
}
\affiliation{
Department of Earth and Space Science, Faculty of Science,
Osaka University, Toyonaka, 560-0043 Osaka, Japan
}
\author{
P. Nordblad
}
\affiliation{
Department of Materials Science, Uppsala University, 
Box 534, SE-751 21 Uppsala, Sweden 
}

\date{\today}

\begin{abstract}
The aging in a Heisenberg-like spin glass \AgMn  is 
investigated by measurements of the zero field cooled magnetic
relaxation at a constant temperature after small temperature shifts $|\Delta T/T_g| < 0.012$. A crossover from fully accumulative to non-accumulative aging is observed, and by converting time scales to length scales using the logarithmic growth law of the droplet model, we find a quantitative evidence that positive and negative temperature 
shifts cause an equivalent restart of aging (rejuvenation) 
in terms of dynamical length scales. This result supports the existence 
of a unique overlap length between a pair of equilibrium states 
in the spin glass system.
\end{abstract}

\maketitle

Spin glasses can be considered as prototype systems for other, more
complex, randomly frustrated and glassy systems \cite{JP}. Experimental
protocols proposed for spin-glasses 
\cite{sg-experiment-review,sg-experiment-review-saclay,jonas},
have later  been employed to study other slowly relaxing systems,
e.g. orientational glasses \cite{doussineau}, polymers \cite{bellon} 
and even interacting nanoparticle systems \cite{nanop}, which all exhibit
qualitatively the same effects of rejuvenation and memory as spin-glasses. 
The modeling of such non-equilibrium phenomena is thus of great
importance for a broad range of physical systems. 
Various scenarios, based on quite different physical ideas, have been 
proposed to account for the empirical behavior, they include 
hierarchical models \cite{sg-experiment-review-saclay,JP,BDHV02,BB02} as well as the 
real space droplet model \cite{droplet,YLB00}.
The simultaneous occurrence of rejuvenation and memory in spin glass dynamics 
has been interpreted to favor the hierarchical scenario \cite{JP}, which does not 
include the temperature chaos concept of the droplet model. On the other hand, it 
was recently found that a standard picture of the droplet model does allow a novel 
dynamical memory due to the existence of 'ghost domains' \cite{YLB00}.

In this Letter we use zero-field-cooled (ZFC) relaxation measurements 
with $T$(temperature)-shifts to investigate rejuvenation in 
the canonical Heisenberg  spin-glass Ag(11 at\% Mn). 
By a new strategy, a dense set of data illustrating the 
crossover from fully accumulative to non-accumulative aging have been obtained.
These results are then analyzed quantitatively within the droplet 
model \cite{droplet} using the concept of temperature-chaos and the recently 
derived parameters of the growth law for the spin glass domains of
Ag(11 at\% Mn) \cite{isothermal}. 
Chaos with temperature is considered as a consequence of the subtle 
competition between fluctuations of the energy and entropy in 
randomly frustrated systems \cite{droplet,SY}. 
According to the droplet theory, 
typical spin configurations of a pair of equilibrium states at 
two different temperatures, say $T_{1}$ and $T_{2}$, are roughly
the same at short length scales much below the so-called overlap 
length $\Lovlp$ but completely different at large length scales 
much beyond $\Lovlp$.  The crossover between the two limits
can be very slow. A remarkable feature is that the overlap length 
is finite for an arbitrary small but non-zero temperature difference 
$\Delta T=T_{1}-T_{2}$.
Consequently, a continuous sequence of re-entrant like phase 
transitions should take place with decreasing temperature throughout the glassy phase. 
 
The ZFC relaxation experiments were performed in a non-commercial squid magnetometer \cite{magetal97}, using the following protocol:
the system is quenched from a temperature above the spin-glass
transition temperature $T_{\rm g}$ ($\approx 33$ K) to an initial temperature 
$T_{i} < T_{\rm g}$, at which it is aged a time $\tw$.
The temperature is then shifted to a measurement temperature $T_{m}=T_{i} + \Delta T$
with $\Delta T$ being either positive or negative. 
Immediately after
reaching temperature stability at $T_{m}$ a weak magnetic field is applied and  
the ZFC magnetization is recorded as a function of time.
For convenience,
let us refer to this $T$-shift experiment as $(T_{i},T_{m})$.
The experiments were performed at temperatures between  
$29.5$ and $30.5$ K and the wait 
times at $T_{i}$ ranged from 100 to 100 000 s. 

\begin{figure}[ht]
\includegraphics[width=0.5\textwidth]{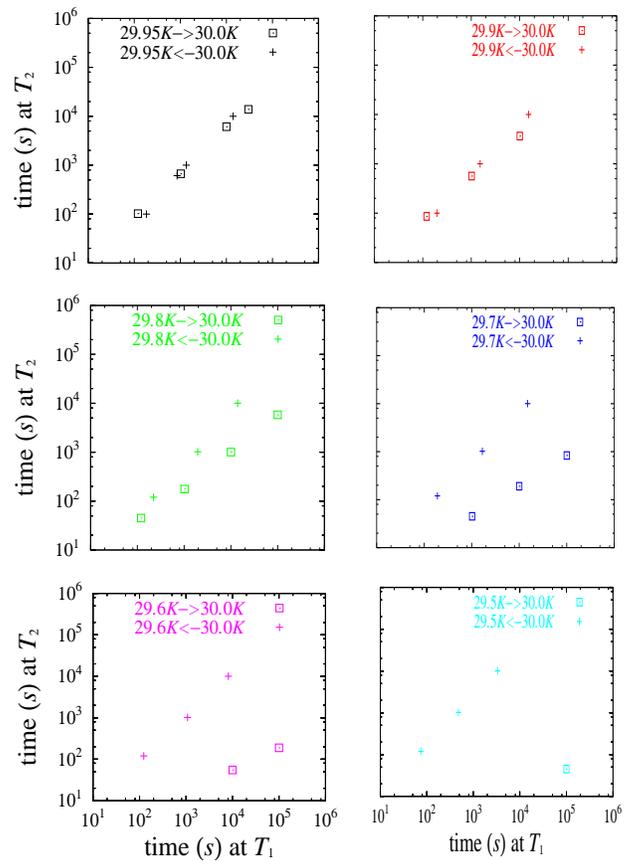}
\caption{Relation between $\tw$ and $\teff$
in twin-experiments - $(T_1,T_2)$ and $(T_2,T_1)$. 
$T_1=30-\Delta T K$ and $T_2=30K$ with $\Delta T$=0.05,
 0.1, 0.2, 0.3, 0.4 and 0.5 K.
The figure demonstrates the appearance of non-accumulative effects.
} 
\label{fig-twin}
\end{figure}

If temperature-chaos is absent, successive aging at the two different 
temperatures will add to each other in a fully accumulative way. 
In such a situation one expects to find an {\it effective age} $\teff$ such that
the ZFC magnetization right after the $T$-shift is already relaxed 
as if it had spent a time $\teff$ by an isothermal aging at $T_{m}$.
The effective age $\teff$ is a monotonically increasing function of $\tw$
and also depends on the temperature protocol $(T_i,T_m)$. Conversely,
$\tw$ can be related to $\teff$ by an inverse function,
\be
\teff=f \left( \tw, (T_i,T_m) \right), \qquad
\tw=f^{-1} \left( \teff, (T_i,T_m) \right),
\ee
It is useful to consider {\it twin-experiments} - $(T_1,T_2)$
and its conjugate $(T_2,T_1)$ for  
a pair of temperatures $T_1$ and $T_2$ below $T_{g}$. 
In the absence of temperature-chaos, there must be a simple 
relation indicating the reversibility of the aging between 
the two temperatures, in such a way that for any time $t$,
\be
f^{-1}\left(t, (T_1,T_2)\right)=f\left(t, (T_2,T_1)\right).
\label{eq-accumulative-relation}
\ee
Let us call it {\it criterion for accumulative aging}.
Any deviation from this simple relation means that
non-accumulative effects exist.  
In the present article we demonstrate evidence for 
a violation of the criterion \eq{eq-accumulative-relation}
(See Fig. \ref{fig-twin})
and show that the scaling ansatz for the temperature-chaos effect 
by the droplet theory can explain the effect quantitatively.

\begin{figure}[ht]
\includegraphics[width=4cm,angle=-90]{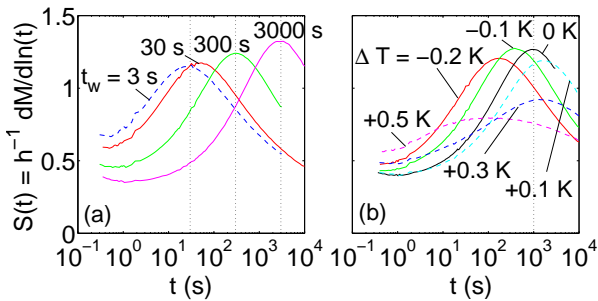}
\caption{Relaxation rate vs time on a logarithmic scale measured at  $T_m=30$ K after (a) aging at $T_m$ for different wait times $\tw$ and (b) aging at $T_m + \Delta T$ for $\tw = 1000$ s.  
\label{S}}
\end{figure}

In an isothermal ZFC relaxation experiment in which the sample has been aged at 
the measurement temperature for a time $\tw$, the relaxation rate 
$S(t)= d M_{\rm ZFC}(t) / d \log(t)$ shows a maximum at $S(t_{\rm max})$ 
with $t_{\rm max} \approx \tw$ \cite{lunetal83}, as illustrated in 
Fig. \ref{S}(a). 
An {\it effective age} $\teff$ after a $T$-shift can accordingly be determined 
from the maximum of $S(t)$ measured after a $T$-shift \cite{sanetal88} (see 
Fig. \ref{S}(b)). 
Each ($T_i,T_m$) experiment is thus associated with the pair of times ($\tw,\teff$). 
However, due to the finite cooling rate in the experiments and the time needed to 
stabilize the temperature, a shortest measurable effective wait time of order 20 s is 
imposed on the system. In the analyzes below, all results from relaxation curves 
yielding $\teff \le$ 30 sec have therefore been discarded. 
As is seen in Fig. \ref{S}(b) there is, for $\Delta T> 0$, a significant broadening of 
the maxima in S(t) compared to the reference S(t) curves. This 
apparent broadening in time scale can, within the droplet scaling 
model, be translated to a dependence of the width of the domain boundaries on the sign 
of the temperature shifts. This subject will be further 
discussed in a forthcoming publication \cite{unp}.

Fig. \ref{fig-twin} shows a set of data from twin-experiments - $(T_1,T_2)$
and $(T_2,T_1)$.  The results of $(T_1,T_2)$ experiments are shown with $\tw$ on 
the horizontal axis and 
$\teff$ on the vertical axis. The results of 
$(T_2,T_1)$ experiments are shown in the reversed way - with $\teff$ on the horizontal axis and $\tw$ on the vertical axis. 
The criterion for accumulative aging \eq{eq-accumulative-relation}
will be satisfied if and only if the two data sets merge with each other
in our plot. It can be seen that for small temperature differences
the data sets merge almost completely which implies a fully
accumulative effect. However for $\Delta T \geq 0.2$K, it is seen
that the criterion is severely violated and
we conclude that a non-accumulative effect is detected.

In order to proceed to a quantitative analysis 
of the non-accumulative effect in terms of the droplet theory, one needs a good 
characterization of the growth law $L_{T}(t)$ for the spin-glass domains
at a given temperature $T$ at time $t$ after the temperature quench.
To this end, we employ the logarithmic domain growth law \cite{droplet,isothermal}
\begin{equation}
L_{T}(t)
\sim L_{0}\left[ \frac{\kb T}{\Delta(T)}
\ln \left(\frac{t}{\tau_{0}(T)}\right) \right]^{1/\psi} \,.
\label{eq-growth} 
\end{equation}
The effects of critical fluctuations are taken into account 
in a renormalized way by the characteristic 
energy scale  $\Delta(T)$ for the free-energy barrier and the characteristic 
time scale $\tau_{0}(T)$ for the thermally activated processes.
They scale as
$\Delta(T)/J= \epsilon^{\psi\nu}$ and
$\tau_{0}(T)/\tau_{\rm m} \sim  (\xi(T)/L_{0})^{z} \sim  |\epsilon|^{-z\nu}$
with  $\epsilon = T/T_g - 1$. The microscopic time scale is
$\tau_{\rm m} \sim \hbar/J \sim 10^{-13}$ s in spin systems, $J \sim T_g$ sets the energy unit, and $z$ and $\nu$ are the dynamical critical exponent and the exponent
for the divergence of the correlation length $\xi(T)$, respectively.
In the following we use $\tau_{m}=10^{-13}$ s, $T_{g}=32.8$ K, $z\nu=7.2$, 
$\psi=1.2$ (from \cite{isothermal}),
$J=T_{g}$ and $\nu=1.1$. The value of $\nu$ 
is known to be around $1.3 \pm 0.2$ \cite{LO86}. 
The length scale that we can explore in the experiments is
much larger than in numerical simulations \cite{kometal2000,PRR01,BB02}
but limited in the range between $L_{T} \sim 130L_{0}$ and 
$L_{T} \sim 180L_{0}$ at our working temperatures.
The advantage of working in terms of dynamical length scale is that it allows 
an understanding of both temperature and time dependencies in a unified way.

Just after a temperature quench (with infinite cooling rate), the spin configuration 
would be 
completely out-of-equilibrium. However,
in the experimental situation the cooling rate is finite (here 3K/min), and the initial
spin configuration when attaining temperature stability at $T{_i}$ has a domain size 
$L_{T_{i}} \gg L_0$ governed by the cooling rate. 
After the aging at the initial temperature $T_{i}$, the projection of the
temporal spin configuration onto the typical spin configuration
of the equilibrium state at $T_{i}$ will have a domain size of order
$L_{T_i}(\tw)$. The spin configuration attained just at the end of the aging 
at $T_{i}$ is essentially the initial spin configuration after the small 
$T$-shift to the measurement temperature $T_{m}$. 
The projection of this configuration onto the typical 
equilibrium spin configuration at $T_{m}$  will
have a certain spatial coherence, $\Leff$.
This length is determined as
\be
\Leff = L_{T_{m}}(\teff),
\label{eq-leff-teff}
\ee
using the growth law Eq.\ (\ref{eq-growth}) quoted above.
In the absence of the temperature-chaos effect, one simply expects that 
aging should be fully accumulative i.e. $\Leff=L_{T_{i}}(\tw)$, which satisfies the
criterion for accumulative aging \eq{eq-accumulative-relation}.
In the presence of temperature-chaos, this is the case 
only at length scales $L_{T_{1}}(\tw) \ll  \Lovlp$, where $ \Lovlp$ is the 
overlap length. $\Leff$ saturates to  
$\Lovlp$ on length scales $L_{T_{i}}(\tw) \gg \Lovlp$. A simple possibility
is then that the two limits are connected by a crossover scaling form,
\be
\frac{\Leff}{\Lovlp}=F\left(\frac{L_{T_{i}}(\tw)}{\Lovlp}\right),
\label{eq-leff}
\ee
with a scaling function $F(x)=x$ for $x \ll 1$ (accumulative aging)
and $F(x)=1$ for $x \gg 1$  (chaos). 
Note that the intermediate regime between the two extremes can be a very
slow crossover. 

In the limit of small temperature differences $|\Delta T/J| \ll 1$,
the overlap length between the two temperatures $T_{1}$
and  $T_{2}=T_{1} + \Delta T$ is supposed to scale as \cite{droplet}
\be
\Lovlp \sim L_{0}|\Delta T/J|^{-1/\zeta} \, , 
\qquad 
\zeta=d_{\rm s}/2-\theta \, ,
\label{Eq: lovlp}
\ee
where $\zeta$ is the chaos exponent. $d_{s}$ is the fractal dimension and $\theta$ the stiffness exponent, which characterizes the surface volume and free-energy gap of the 
droplets, respectively.

\begin{figure}[t]
\includegraphics[width=0.5\textwidth]{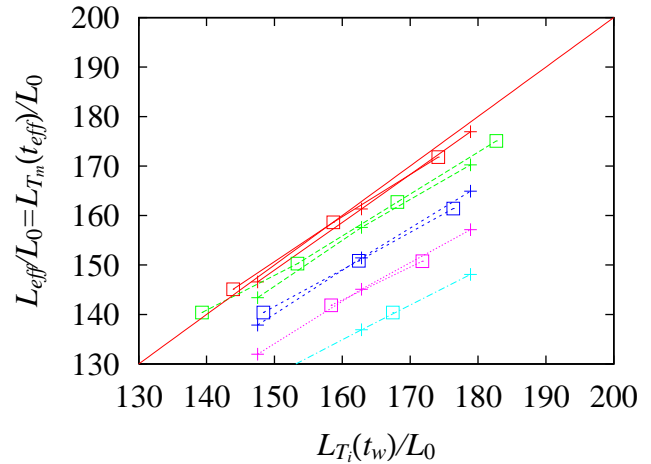}
\caption{Effective domain size after $T$-shifts. 
Data from the twin-experiments - $(T_1,T_2)$ ,$(T_2,T_1)$ with $T_1=30 K -\Delta T$ and 
$T_2=30 K$ with $\Delta T$=0.1, 0.2, 0.3, 0.4 and 0.5 K (from top to bottom).
%For a given pair- $T_1$ and $T2$ -the same
%color is used. 
$(T_1,T_2)$ experiments are indicated by square markers and $(T_2,T_1)$ by crosses. 
The solid straight line 
represents the case of fully accumulative aging.  This plot demonstrates
equivalence between positive and negative $T$-shifts.}
\label{Fig: lteff}
\end{figure}

\begin{figure}[t]
\includegraphics[width=0.5\textwidth]{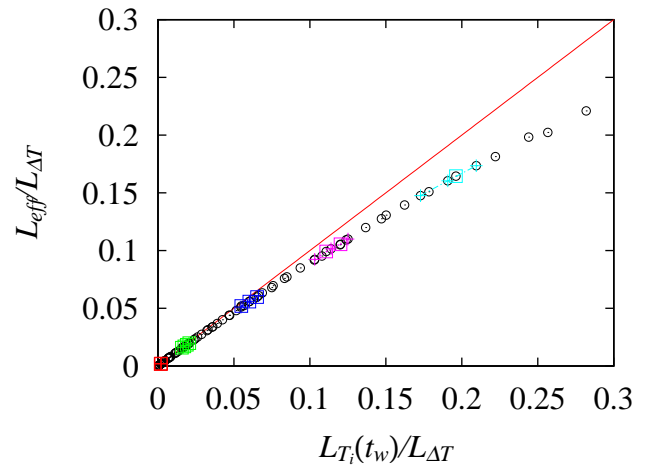}
\caption{Scaling plot of $\Leff$ with $\Lovlp= L_0 (c \Delta T/J)^{-2.6}$ 
with $c=5$. The solid straight line 
represents the case of fully accumulative aging.
The open circles mark the additional data.}
\label{Fig: scale-lteff}
\end{figure}

In Fig.\ \ref{Fig: lteff}, we present the results of the 
twin-experiments - $(T_1,T_2)$ ,$(T_2,T_1)$ - converting $\teff$ and $\tw$ into 
the domain sizes $\Leff$ and $L_{T_i}(\tw)$ using
the growth law \eq{eq-growth} and \eq{eq-leff-teff}. It is seen that the data 
fulfilling the criterion for accumulative aging in Fig. \ref{fig-twin} lie on the 
straight line corresponding to $\Leff = L_{T_i}(\tw)$ and
equally importantly that the data of the $(T_1,T_2)$ and $(T_2,T_1)$ experiments merge
with each other. The latter clearly demonstrates 
that the effective domain size for a given pair of temperatures is the same which 
is consistent with the expectation that there is a unique
overlap length between a given pair of temperatures. 

Furthermore, the data with $\Delta T > 0.1 $K shown in Fig. \ref{Fig: lteff} lie below
the straight line corresponding to $\Leff=L_{T_i}(\tw)$, which indicates a 
finite overlap length, associated with the non-accumulative effects already observed in 
Fig. \ref{fig-twin}. 
Fig. \ref{Fig: scale-lteff} shows a scaling plot 
to fully test \eq{eq-leff} using the data shown in Fig. \ref{Fig: lteff} 
and an additional set of data for a variety of $\Delta T$s. 
A good collapse of the
data is obtained for $1/\zeta=2.6\pm 0.5$. It was found in Ref. \cite{isothermal}
that $\theta \sim 1.0$ in the present sample. Then the scaling 
relation \eq{Eq: lovlp} implies $d_{s}=2.8$ which 
is consistent with the boundaries \cite{droplet} $d-1 < d_{s} <d$ with $d=3$. 

In the above discussion we have concentrated on one particular feature 
in the relaxation rate curves, the position of the maximum, $\teff$. 
The observation time where this occurs is then understood to reflect 
an effective domain size $\Leff$ in the spin glass. 
Looking at the full relaxation rate curves there are of course other
features that give additional information such as about the progressive change of the 
population of thermally active droplets in the interior of the domains 
which should enhance the response. Such additional
features could also obscure the determination of $\teff$. An alternative method is 
then to derive $\teff$ from 
the out-of phase component of the ac-susceptibility as in previous studies 
of the effect $T$-shifts \cite{kometal2000,saclay,BB02}.
We have performed some preliminary measurements on the Ag(Mn) sample \cite{unp} 
using the method of \cite{kometal2000,saclay,BB02,takhuk} following 
identical thermal protocols as in the current study and found that the obtained 
values of $\teff$ are consistent in-between the two methods.

%The $\teff$ values derived by this method are found to be smaller than the ones derived from the relaxation rate and to depend on frequency. 
%This inconsistency problem and possible explanations will be discussed in detail elsewhere.

In this article we have limited ourselves to investigate the small $\Delta T$ case
in order to elucidate universal features of the temperature-chaos effect.
Although the departure from the fully accumulative limit
is modest in terms of length scales as shown 
in Fig. \ref{Fig: scale-lteff}, the impact 
is already striking in terms of time scales as seen in Fig. \ref{fig-twin}.

In experiments most systems appear less ``chaotic'' 
than the canonical Ag(Mn) and Cu(Mn) spin glasses 
\cite{saclay,roland,bellon,doussineau,nanop}. 
An open question is if this is due to the absence 
of chaos {\it or} only a smaller  $1/\zeta$, 
e.g. for Ising spin glasses $\theta \approx 0.2$ 
imposes $1/\zeta \approx 1$ compared to 
$1/\zeta \approx 2.6$ for the Ag(Mn) spin glass. 
Hence, the droplet theory implies that
Ising spin glasses should appear 'less' chaotic than Heisenberg 
spin glasses due the smaller chaos exponent (i.e. larger $\Lovlp$). 

In summary,  we have shown that results from $T$-shift experiments 
on the dynamic susceptibility of 
an archetypical spin glass when 
parametrized into a dynamical length scale 
(Fig. \ref{Fig: lteff}) yield an anticipated equivalence between
positive and negative $T$-shifts and support the concepts of overlap length 
and temperature-chaos. 
The possibility to scale all data onto one master curve (Fig. \ref{Fig: scale-lteff})
further confirms the applicability of the employed spin glass domain model 
\cite{droplet}.
We hope that the results of the present paper will stimulate further quantitative 
analyzes of rejuvenation-memory effects in
spin-glasses and other slowly relaxing systems in terms of real space as well as 
hierarchical models to reach a consistent modeling of aging dynamics.

\acknowledgments
We would like to thank Hajime Takayama, Koji Hukushima and Roland Mathieu for stimulating discussions.
This work was financially supported by the Swedish Research Council (VR).

\end{document}